\documentclass[twocolumn,aps,floatfix,showpacs]{revtex4}
\newcommand{\be}{\begin{equation}}
\newcommand{\ee}{\end{equation}}
\newcommand{\bea}{\begin{eqnarray}}
\newcommand{\eea}{\end{eqnarray}}

\usepackage{graphicx}
\usepackage{bm}
\begin{document}

\title{Feshbach spectroscopy of a K-Rb atomic mixture}
\author{Francesca Ferlaino, Chiara D'Errico, Giacomo Roati, Matteo Zaccanti,
Massimo Inguscio, and Giovanni Modugno}
\affiliation{LENS and
Dipartimento di Fisica, Universit\`a di Firenze,
  and INFM-CNR\\  Via Nello Carrara 1, 50019 Sesto Fiorentino, Italy }

\author{Andrea Simoni}
\affiliation{Laboratoire de Physique des
Atomes, Lasers, Mol\'ecules et Surfaces \\ UMR 6627 du CNRS and
Universit\'e de Rennes, 35042 Rennes Cedex, France}

\begin{abstract}
We perform extensive magnetic Feshbach spectroscopy of an ultracold
mixture of fermionic $^{40}$K and bosonic $^{87}$Rb atoms. The
magnetic-field locations of 14 interspecies resonances is used to
construct a quantum collision model able to predict accurate
collisional parameters for all K-Rb isotopic pairs. In particular we
determine the interspecies $s$-wave singlet and triplet scattering
lengths for the $^{40}$K-$^{87}$Rb mixture as $(-111\pm 5)a_0$ and
$(-215\pm 10)a_0$ respectively. We also predict accurate scattering
lengths and position of Feshbach resonances for the other K-Rb
isotopic pairs. We discuss the consequences of our results for
current and future experiments with ultracold K-Rb mixtures.

\end{abstract}

\date{\today}
\maketitle

Quantum degenerate atomic mixtures
\cite{mixtures1,mixtures2,mixtures3,roati,modugno,mixtures4} are
promising for the study of a variety of novel physical phenomena,
such as production of quantum gases of polar molecules \cite{polar},
boson-induced superfluidity of fermions \cite{pairing}, and quantum
phases of matter in optical lattices \cite{lattices} or random
potentials \cite{random}. The study of these phenomena requires the
use of magnetically tunable Feshbach resonances \cite{inouye} to
control the interaction. A detailed knowledge of resonances and
collisional parameters is however necessary to achieve such control.
Feshbach resonances have been deeply investigated in homonuclear
systems, and recently observed also in some heteronuclear mixtures
\cite{mit,jila,ens}. However, accurate Feshbach spectroscopy, has
been performed so far only on homonuclear systems
\cite{cs,munich,stuttgart}. We report here an extensive experimental
study of Feshbach resonances in a $^{40}$K-$^{87}$Rb mixture. This
is used to construct a quantum collisional model able to predict the
relevant parameters for all K-Rb isotopic pairs, including both
boson-fermion and boson-boson pairs of great experimental interest.

In particular this study allows us to determine univocally the
scattering lengths for the $^{40}$K-$^{87}$Rb mixture, for which
contrasting determinations have been reported. The values we find
for the singlet $a_s$ and triplet $a_t$ scattering lengths are
$(-111\pm5)a_0$ and $(-215\pm10)a_0$, respectively. We also
determine high-accuracy values for inter-isotope triplet and singlet
scattering lengths for all other K-Rb pairs. In addition, we
determine all relevant parameters of Feshbach resonances in the
$^{40}$K-$^{87}$Rb mixture and demonstrate the presence of several
Feshbach resonances in other K-Rb isotopic pairs. Our findings have
important consequences for both ongoing and future experiments with
K-Rb mixtures.

The apparatus and techniques used have already been presented in
detail elsewhere \cite{roati} and are only briefly summarized here.
We prepare samples of typically 10$^5$ K fermions and
5$\times$10$^5$ Rb bosons at ultralow temperatures using two
successive phases of laser and evaporative cooling. The mixture is
magnetically trapped in the states $|F=9/2, m_F=9/2\rangle$ for K
and $|2, 2\rangle$ for Rb. The atomic sample is then adiabatically
transferred to a purely optical trap, created by two off-resonance
laser beams, at a wavelength of 830~nm, crossing in the horizontal
plane. The typical density of the Rb (K) sample in the optical trap
throughout this experiment was 5$\times$10$^{12}$~cm$^{-3}$
(1$\times$10$^{12}$~cm$^{-3}$), while the common temperature of the
two gases was about 1~$\mu$K. At this temperature the two samples
are still out of the quantum regime.

\begin{figure}[htbp]
\includegraphics[width=\columnwidth,clip]{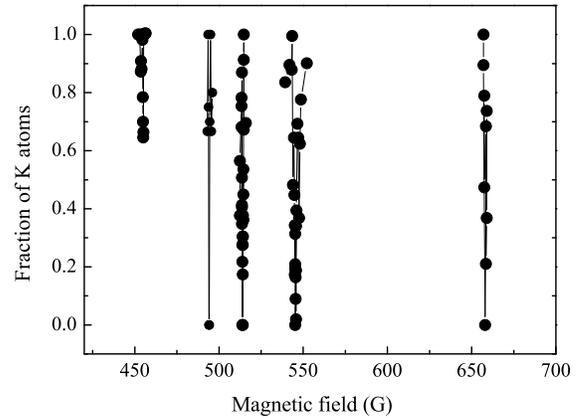}
\caption{Relative inelastic losses of potassium atoms in a
$^{40}$K-$^{87}$Rb mixture in its absolute ground state at
interspecies Feshbach resonances. The two features near 456~G and
515~G are $p$-wave resonances, the others are $s$-wave resonances.}
\label{fig1}
\end{figure}

The atoms are then transferred in selected states by means of a
series of radio-frequency (RF) and microwave ($\mu$w) adiabatic
rapid passages. As already noted in \cite{simoni} any combination of
states in which one species is in its absolute ground state and the
other is in any magnetic sublevel of its ground hyperfine state, is
stable against spin-exchange collisions. In this work we have
studied in particular the combinations with Rb in $|1, 1\rangle$ and
K in: a)$|9/2, -9/2\rangle$; b)$|9/2, -7/2\rangle$; c) $|9/2,
7/2\rangle$. In a first phase we apply a homogeneous magnetic field
of about 10~G and then transfer Rb to the $|1, 1\rangle$ state and K
to the $|9/2, -9/2\rangle$ (for cases a and b) by means $\mu$w and
RF sweeps, respectively. The field is then raised to 100~G to
perform the additional K transfer $|9/2, -9/2\rangle\rightarrow
|9/2, -7/2\rangle$ ($|9/2, 9/2\rangle\rightarrow |9/2, 7/2\rangle$)
for case b(c) with another RF sweep. The homogeneous field is then
changed in few ms to any value in the range 0-900~G and actively
stabilized there with an accuracy of about 200~mG. The field is
calibrated by means of RF spectroscopy of the $|2, 2\rangle$ - $|2,
1\rangle$ transition of Rb.

To look for Feshbach resonances we record the fraction of K atoms
lost through inelastic collisions after about 1~s of permanence in
the magnetic field. The typical experimental signature of
interspecies resonance is shown in Fig.~\ref{fig1} for the
absolute ground state of the mixture ($|9/2, -9/2\rangle$+$|1,
1\rangle$): due to the lower abundance of K in our sample, we
usually observe a complete loss of K atoms at resonance. To avoid
any possible confusion with homonuclear resonances we also check
the absence of losses after removing either K or Rb from the
mixture before applying the magnetic field.

The first theoretical study on this system \cite{simoni} predicted
the presence of several resonances around 500~G for most hyperfine
states. In this experiment we have detected 14 of such resonances,
including the four in the ground state already investigated in
Ref.~\cite{jila}. The measured positions and widths are summarized
in Tab.~\ref{table1}. While our measurements confirm the
magnetic-field positions of the four resonances reported in
~\cite{jila}, most of the newly observed resonance pattern cannot be
reproduced using the collision model proposed
in~\cite{jila,goldwint}, indicating the need for an alternative
interpretation.

Our theoretical quantum collision model is constructed as described
in Ref.~\cite{simoni,ferrari}. The isotropic singlet $X^1\Sigma^+$
and triplet $a\Sigma^+$ interaction potentials are parameterized in
terms of the $a_s$ and $a_t$ scattering lengths, respectively. Few
sample experimental features are compared to maxima in the two-body
elastic cross section computed for different $a_{s,t}$, until a good
agreement is found. A global least square fit is then performed,
leading to the best fit parameters with one standard deviation
$a_s=(-111\pm5)a_0$ and $a_t=(-215\pm10)a_0$. In the fit also the
van der Waals coefficient $C_6$ is let free to vary, obtaining
$C_6=(4292\pm 19)$~a.u., which agrees to better than one standard
deviation with the high precision {\it ab-initio} calculation of
\cite{derevianko}. Error bars also include a typical $\pm 10\%$
uncertainty in $C_8$. The average theory-experiment deviation for
the resonance positions is about 0.3~G only.

\begin{table}[t]
\begin{center}
\caption{Magnetic-field positions and widths of the observed
$^{40}$K-$^{87}$Rb resonances compared to the corresponding
theoretical predictions of our best-fit model (see text).
$\Delta_{\rm exp}$ is defined as the 1/e$^2$ halfwidth of a gaussian
best fit to the atom loss profile, while $\Delta_{\rm th}$ is
defined in the text, and $\ell$ is the orbital angular momentum of
the molecular state associated to each resonance.} \label{table1}
\vskip 12pt
\begin{tabular}{l | c c c cc}
\hline \hline
 $|m_{fa}\rangle$+$|m_{fb}\rangle$& $B_{\rm exp}$ (G) & $\Delta_{\rm exp}$(G)  & $B_{\rm th}$(G)&$-\Delta_{\rm th}$ (G)& $l$ \\
 \hline \hline
$|-9/2\rangle$+$|1\rangle$
& 456.0 & 0.2& 456.5 &$2~10^{-3}$&1\\
& -- & -- & 462.2   & 0.067  &0\\
&495.6 & 0.5& 495.7 &0.16 &0\\
&515.7 &0.5&515.4 & 0.25  &1\\
&546.7&1.2&546.8&2.9&0\\
&658.9&0.6&659.2& 1.0 &0\\
&663.7&0.2&663.9& 0.018 &2\\
\hline
$|-7/2\rangle$+$|1\rangle$& 469.2 &0.4&469.2&0.27&0\\
&--&--&521.6&0.051&0 \\
&584.0&1.2&584.1&0.67&0 \\
& 591.0 &0.3&591.0 &$2~10^{-3}$&2 \\
& 598.3 & 0.6&598.2&2.5&0\\
& 697.3 &0.3&697.3 &0.16&0 \\
& 705.0 &1.4&704.5 &0.78&0\\
\hline $|7/2\rangle$+$|1\rangle$&299.1&0.3&298.6&0.59&0\\
& 852.4 & 0.8& 852.1&0.065& 0\\
\hline \hline

\end{tabular}
\end{center}
\end{table}

The nature of the molecular states associated to the resonances can
be better understood through multichannel bound state calculations.
Since $a_s$ and $a_t$ are comparable the spacing between singlet and
triplet vibrational levels is small compared to the hyperfine
interaction.  Strong singlet/triplet mixing then occurs at least for
the two vibrational states closest to dissociation, resulting in
molecular levels labeled as $(F_a F_b F \ell)$ in {\it zero}
magnetic field, where $\ell$ is the rotational quantum number and
$\vec F = \vec{F}_a +\vec{F}_b$. The features below $\approx600$~G
arise from these strongly mixed levels. At such magnetic fields
however the Zeeman magnetic energy is comparable to the smaller
hyperfine splitting in the system, that of $^{40}$K. Therefore $F_a$
is not a good quantum number to label the resonances, whereas $F_b$
is approximately good and equal to 2. Resonances at higher magnetic
field correlate with more deeply-bound states and tend to assume
singlet or triplet character. In all cases $\ell$ is an almost exact
quantum number and is also shown in Tab.~\ref{table1}.

In Tab.~\ref{table1} we also show the theoretical width $\Delta_{\rm
th}$ of the resonances, defined as the difference between the
magnetic-field locations of maximum and minimum of the elastic cross
section. The $\ell$=0 molecules tend to give rise to broad
resonances due to strong spin-exchange coupling to incoming $s$-wave
atoms. We also observe two narrow resonances due to coupling of a
$\ell$=2 molecule to incoming $s$-wave atoms through weaker
anisotropic spin-spin interactions~\cite{cs,stuttgart}. The two
resonances associated with $\ell$=1 molecules couple by
spin-exchange to incoming $p$-wave atoms. These resonances have an
energy-dependent width \cite{bohnp} which we compute at collision
energy $E/k_B \approx 1 \mu$K, with $k_B$ the Boltzmann constant.
The $p$-wave nature of the ground-state resonance near 455~G is
confirmed by its weak relative strength with respect to the
neighboring ones, as shown in Fig.~\ref{fig1} . Conversely, this
relative suppression  was not observed in the experiment reported in
\cite{jila,goldwint}, which was performed at the $\sim 10$ times
larger temperature. Such a decrease of the resonance strength with
the temperature is indeed expected for $p$-wave resonances
\cite{bohnp}. The $\Delta_{\rm exp}$ of the highest-field $p$-wave
feature is larger, as expected because of its larger $\Delta_{\rm
th}$. Tab.~\ref{table1} also shows two narrow not yet observed
resonances. We find that several stable states of the mixture
present at least one broad resonance, analogous to that in the
ground state near 545~G. Any of these resonances can be very well
suited for control of the interaction and molecule formation.

The optimized $^{40}$K-$^{87}$Rb model can now be used to determine
singlet and triplet scattering lengths $\tilde{a}_{s,t}$ for any
K-Rb isotopic pair, see Tab.~\ref{table2}. Within the
Born-Oppenheimer approximation this can be simply achieved by using
the appropriate reduced mass in the Hamiltonian. We note that such
mass-scaling procedure depends in a sensitive way on the actual
number of bound states supported by the potentials. They are
nominally $N_b^{s}=98$ and $N_b^{t}=32$ for the singlet and triplet
{\it ab initio} potentials we use, with an expected uncertainty of
$\pm 2$~\cite{zemke}. In fact, we find that the error in
$\tilde{a}_{s,t} $ due to variation $\delta N_b^{s,t}$ (for fixed
$a_{s,t}$) dominates that due to $a_{s,t}$ (for fixed $N_b^{s,t}$)
for all the isotopic combinations. This shift can be expressed to a
few percent accuracy through \bea \frac{2}{\pi}\,
\delta\arctan{\frac{\tilde{a}_{s,t}}{L}} = \beta_{s,t} \delta
N_b^{s,t} \label{eq1} \eea where $L=72~a_0\approx \frac{1}{2} (2 \mu
C_6)^{1/4}$ is the typical length scale of a van der Waals
potential, with $\mu$ the K-Rb reduced mass. The value of the
K$^{41}$-Rb$^{87}$ triplet scattering length in Tab.~\ref{table2}
confirms the direct collisional measurements reported in
Ref.~\cite{ferrari}. The comparison is not conclusive about the
number of bound states, though an optimal agreement is found for our
nominal $N_b^{t}$. However, a limited amount of Feshbach
spectroscopy on a different pair might be sufficient to determine
$N_b^{s,t}$.

Another quantity of general interest for experiments with K-Rb
mixtures is the effective elastic scattering length $a$ for the
absolute ground state. The $a$ for all isotopes are reported in
Tab.~\ref{table3} together with the location of Feshbach
resonances for the systems we judge most interesting for future
experiments, the three boson-boson pairs $^{39}$K-$^{87}$Rb,
$^{41}$K-$^{85}$Rb, and $^{41}$K-$^{87}$Rb.


\begin{table}[t]
\begin{center}
\caption{Calculated singlet and triplet $s$-wave scattering
lengths for collisions between K and Rb isotopes. Sensitivity
parameters $\beta_{s,t}$ to the number of bound states
(Eq.~\ref{eq1}) are also shown, with power of ten displayed in
parenthesis. } \label{table2} \vskip 12pt

\begin{tabular*}{8cm}{@{\extracolsep{\fill}} c | c c c c }
\hline\hline
 K-Rb  &$\tilde{a}_s$ ($a_0$)&$\beta_s$ & $\tilde{a}_t$ ($a_0$) & $\beta_t$ \\
\hline
 39-85& $26.5\pm0.9$&-2.8(-2)& $63.0\pm0.5$ &-1.3(-2)\\
 39-87&  $824_{-70}^{+90}$   &-1.5(-2)&$35.9\pm0.7$  &-1.6(-2) \\
 40-85& $64.5\pm0.6$ &-3.9(-3)&$-28.4\pm1.6$ &-1.8(-2)\\
 40-87& $-111\pm5$   &        &$-215\pm10$    & \\
 41-85&$106.0\pm0.8$&3.6(-3) &$348\pm10$     &6.5(-3)\\
 41-87& $14.0\pm1.1$ &2.6(-2) &$163.7\pm1.6$ &7.7(-3)\\
\hline\hline
\end{tabular*}
\end{center}
\end{table}

Let us now discuss the results presented so far, beginning with
the $^{40}$K-$^{87}$Rb system for which a comparison with existing
determination of the scattering lengths is due. The values of
$a_{s,t}$ determined here differ from the results
$a_s=(-54\pm12)a_0$ and $a_t=(-281\pm15)a_0$ of the analysis in
Ref.~\cite{jila}, which was based on an incorrect resonance
assignment. The present $a_s$ value is consistent with
Ref.~\cite{simoni}. The $a_t$ is consistent with the value of
Ref.~\cite{ferrari}, which however bears large error bars, and it
is in reasonable agreement with~Ref.\cite{goldwin}. It is
otherwise neither consistent with the determination
of~\cite{damping} nor with the observation of a collapse
instability in this mixture \cite{collapse,hamburg}. Comparison of
the collapse observations reported so far with current mean-field
models indicated $a_t=(-395\pm15)a_0$ for \cite{collapse} (see the
analysis in \cite{modugno}) and (-281$\pm$15)$a_0$ for
\cite{hamburg}. These discrepancies with the current determination
could be in principle explained by a low-field Feshbach resonance
in the magnetically trappable state $|9/2, 9/2\rangle$+$|2,
2\rangle$ used in those experiments, which however seems to be
excluded by our collisional model. Further investigation is
therefore necessary in order to understand the observed
phenomenology. In particular, use of Feshbach resonances to
control the effective interaction should allow the current
theories of instabilities to be tested.

The Fermi-Bose system is also interesting to study boson-induced
fermion pairing \cite{pairing}, due to its large and attractive
background interaction. However, the new value of the ground state
scattering length is considerably smaller than the one considered in
Ref.~\cite{simoni} and the optimal conditions for $s$-wave pairing
described therein seem to be difficult to reach. Moreover, the
present investigation shows that no overlapping resonances which
could be used to favor the pairing \cite{simoni} exist in collisions
between the two lower Zeeman states of K and the ground state of Rb.
The most promising direction seems therefore to be the $p$-wave
boson-induced pairing of fermions \cite{ppairing} at an interspecies
$p$-wave Feshbach resonance.

\begin{table}[t]
\begin{center}
\caption{Predicted zero-field $s$-wave scattering lengths for the
absolute ground state of K-Rb isotopes.  Feshbach resonance
positions and widths are also provided for three selected isotopic
pairs. The quoted uncertainties do not include the uncertainty on
the number of bound states.} \label{table3} \vskip 12pt
\begin{tabular*}{7cm}{@{\extracolsep{\fill}}l@{\hspace{0.5cm}}|c@{\hspace{0.5cm}}|cc}
\hline \hline
K-Rb & $a$ ($a_0$) & $B_{\rm th}$~(G)&$ \Delta_{\rm th}$~(G)  \\
\hline \hline
39-85&$56.6\pm0.4$&&\\
\hline 39-87&$27.9\pm0.9$&$248.8\pm1.6$&0.26\\
&&$320.1\pm1.6$&7.9 \\
&&$531.9\pm1.2$& 2.7 \\
&&$616.2\pm1.5$& 0.10 \\
\hline
40-85&$-21.3\pm1.6$&&\\
\hline
40-87&$-185\pm7$&&\\
\hline
41-85&$283\pm6$&$132.5\pm0.6$&0.19\\
&&$141.2\pm1.1$&$\quad 2~10^{-4}$\\
&&$147\pm2$&0.025\\
&&$184.6\pm1.0$&2.9 \\
&&$191.4\pm1.0$&0.81\\
&&$660\pm3$&3.4\\
&&$687\pm2$&16\\
\hline
41-87&$1667_{-406}^{+790}$&$17\pm5$&45\\
&&$67\pm3$&8.9\\
&&$516\pm7$&82\\
&&$688\pm8$&0.059\\
 \hline \hline

\end{tabular*}
\end{center}
\end{table}

Concerning the other isotopic pairs, the simultaneous determination
of scattering lengths and Feshbach resonances is of invaluable help
in devising future experiments. For example, sympathetic cooling of
$^{39}$K using $^{87}$Rb is interesting to create a Bose-Einstein
condensate with zero-field negative scattering length in the ground
state and a rather broad Feshbach resonance available at low field
in the $|1, -1\rangle$ hyperfine state \cite{bohn}. In spite of a
rather small interspecies scattering length the broad resonance near
320~G predicted in this work could be used in order to enhance
thermalization between the two components. A sample of $^{41}$K
could instead be used to optimize the evaporation of $^{85}$Rb atoms
which is typically very inefficient in a pure homonuclear sample due
to occurrence of the first zero in the $^{85}$Rb cross-section
already at temperatures on the order of 100~$\mu$K \cite{wieman}. On
the converse, we have checked that the large magnitude of the zero
energy interspecies cross section persists even up to the mK regime.
The availability of several Feshbach resonances at relatively low
field could also prove to be useful for the production of binary
Bose-Einstein condensates where both the self- and the interspecies
interaction are tunable. Finally, in the $^{41}$K-$^{87}$Rb pair, a
system for which the production of a stable binary condensate has
already been reported \cite{modugno}, availability of very broad
resonances will allow the mutual interaction to be precisely tuned.
The availability of heteronuclear resonances in these mixtures could
also be exploited for the formation of bosonic polar molecules.

In conclusion, we have performed extensive Feshbach spectroscopy in
the heteronuclear $^{40}$K-$^{87}$Rb system and constructed an
accurate collisional model capable of predicting scattering lengths
and Feshbach resonances for all K-Rb isotopic pairs. This will serve
as invaluable input to future experiments on these mixtures. In
particular, our accurate characterization of Feshbach resonances in
K-Rb mixtures will be useful to investigate the formation of both
fermionic and bosonic heteronuclear molecules.

We acknowledge participation of E. de Mirandes and H. Ott to the
early stage of the experiment and contributions by T. Ban, V. Josse,
and R. Salem. We thank all the LENS Quantum Gases group for useful
discussions. This work was supported by MIUR, by EU under contracts
HPRICT1999-00111 and MEIF-CT-2004-009939, and by Ente CRF, Firenze.

\end{document}